\documentclass[twocolumn,aps,pra,eqsecnum]{revtex4-1}
\usepackage[T1]{fontenc}
\usepackage[latin9]{inputenc}
\usepackage{color}
\usepackage{bm}
\usepackage{amsmath}
\usepackage{amssymb}
\usepackage{graphicx}
\usepackage{esint}

\makeatletter
 
 \@ifundefined{textcolor}{}
 {%
   \definecolor{BLACK}{gray}{0}
   \definecolor{WHITE}{gray}{1}
   \definecolor{RED}{rgb}{1,0,0}
   \definecolor{GREEN}{rgb}{0,1,0}
   \definecolor{BLUE}{rgb}{0,0,1}
   \definecolor{CYAN}{cmyk}{1,0,0,0}
   \definecolor{MAGENTA}{cmyk}{0,1,0,0}
   \definecolor{YELLOW}{cmyk}{0,0,1,0}
 }

\@ifundefined{definecolor}
 {\usepackage{color}}{}
\@ifundefined{definecolor}{\usepackage{color}}{}

\usepackage[english]{babel}

\makeatother

\begin{document}

\title{Quantum simulations of the early universe}

\author{B. Opanchuk$^{1}$, R. Polkinghorne$^{1}$, O. Fialko$^{2}$, J.
Brand$^{2}$, P. D. Drummond$^{1}$}

\email{pdrummond@swin.edu.au}

\selectlanguage{english}%

\affiliation{$^{1}$Centre for Atom Optics and Ultrafast Spectroscopy, Swinburne
University of Technology, Melbourne 3122, Australia}

\affiliation{$^{2}$Institute of Advanced Studies, Massey University, Albany,
New Zealand}
\begin{abstract}
A procedure is described whereby a linearly coupled spinor Bose condensate
can be used as a physically accessible quantum simulator of the early
universe. In particular, an experiment to generate an analog of an
unstable vacuum in a relativistic scalar field theory is proposed.
\textcolor{black}{This is related to quantum theories of the inflationary
phase of the early universe.} There is an unstable vacuum sector whose
dynamics correspond to the quantum sine-Gordon equations in one, two
or three space dimensions. Numerical simulations of the expected behavior
are reported using a truncated Wigner phase-space method, giving evidence
for the dynamical formation of complex spatial clusters. Preliminary
results showing the dependence on coupling strength, condensate size
and dimensionality are obtained.
\end{abstract}
\maketitle

\section{Introduction}

The use of quantum simulation as a route to better understanding of
complex quantum dynamics has much to recommend it. Conceptually, this
creates an analog quantum computer. In quantum simulations, a table-top
experiment is carried out to mimic a more complex quantum system which
we would like to understand. This method has been used to treat, for
example, models of condensed matter phase-transitions~\cite{Porras2004}.
Such an approach can be complemented by the use of approximate computer
simulations. A numerical simulation can then be verified and tested
in the table-top experiment, while allowing a wider variety of parameters
to be treated.

But why stop at condensed matter: why not model the entire universe?
This seems presumptuous, and possibly is. The entire universe will
never be shoehorned into a table-top experiment with every complexity
intact. Nevertheless, cosmologists today often use quantum field theory
models to describe the early universe. This can be traced back to
the pioneering works of Higgs and colleagues~\cite{Higgs1964,Englert1964,Higgs1964a,Guralnik1964},
studying the origins of mass, and to Coleman's groundbreaking studies
on unstable quantum vacuum decay~\cite{Coleman1977,Callan1977}.

A combination of these approaches, together with the inclusion of
general relativity, leads to the current inflationary universe scenario~\cite{Guth1981,Linde1991,Linde1994a,Linde1998,Stewart1993,Liddle1994}.
\textcolor{black}{Inflation provides a possible explanation for the
origin of structure in the universe.}\textcolor{red}{{} }Can we model
at least part of this picture of the early universe in a laboratory?
The simplest model for the scalar inflaton field $\phi(x)$ is described
by the Lagrangian

\begin{equation}
{\cal L}=\frac{1}{2}\partial_{\mu}\phi\partial^{\mu}\phi-V(\phi),
\end{equation}
 where $V(\phi)$ is the potential down which the scalar field rolls
or decays. In this paper, we show how to realize the above relativistic
scalar field model in a coupled Bose condensate, and model the fate
of the scalar field of the early universe. We note that there have
been earlier proposals using somewhat different, albeit related techniques~\cite{Fischer2004,Neuenhahn2012}.

\textcolor{black}{While the potential $V(\phi)$ is specific, depending
on the model of inflation under consideration, coupled Bose condensates
provide a situation with $V(\phi)\propto-\cos(\phi)$. In the spirit
of inflation, we consider a cold, unstable vacuum with $\phi=\pi$
a}s the initial condition. The vacuum gradually decays~\cite{Linde1991,Linde1994a,Linde1998},
to produce a hot universe, replete with dynamical clumping into random
structures. This quantum dynamical \emph{``universe on a table-top''}
experiment is probed by means of an internal Rabi rotation, and imaged.

Relativistic quantum field theory is usually tested experimentally
at large accelerators. However, energies at high energy particle accelerators
like CERN are not high enough for these field theories, although observational
evidence for the Higgs boson~\cite{ATLASCollaboration2012,CMSCollaboration2012}
is thought to provide evidence for the low energy sector of scalar
quantum fields in the universe. Instead of accelerating particles
to light speed, we propose, essentially, to slow the velocity of light
down to atomic speeds. This allows us to study novel phenomena outside
of CERN's limits. A quantum simulation of interacting relativistic
fields in two, three or four space-time dimensions has the useful
feature that it allows us to access quantum physics that we cannot
do experiments on by any other means.

We note some recent experiments have explored relevant physics with
ultra-cold atoms, demonstrating that our proposal is indeed feasible.
These show that the physics of interest here is very close to realization.
The investigations that are similar to our proposal include long time-scale
interferometry with a two-level BEC~\cite{Egorov2011}, and the study
of the thermalization of a BEC~\cite{Gring2012}. More recent experiments
have realized a flat trap or ``can'' for a BEC in three dimensions~\cite{Gaunt2013}.
In principle, simulations of this type might eventually produce results
that can predict the early universe temperature maps~\cite{PlanckCollaboration2013}
produced recently by the Planck telescope surveys, or very large-scale
density inhomogeneities. Here we have a more cautious goal, of simply
being able to demonstrate quantum simulations of an unstable quantum
vacuum in a relativistic quantum field theory.

Our starting point is a trapped Bose condensate with two internal
quantum levels. The levels are coupled by a microwave field, and the
atoms interact through S-wave scattering. We assume zero temperatures,
homogeneous couplings and perfect microwave phase stability. While
departures from such perfection can be treated, and these requirements
can be relaxed, we will study the ideal case here. We show that by
introducing a simple $\pi$ phase shift into the coupling field, it
is possible to generate an experimentally accessible model of an unstable
relativistic vacuum. A quantum simulation of the decay of this unstable
vacuum then provides the simplest of early universe models. The relevant
field variable is the relative phase of the two condensates. In a
regime of small coupling, the relative phase obeys the sine-Gordon
equation~\cite{Kaurov2006,Gritsev2007}, a popular model for a relativistic
scalar field theory where the field potential $V(\phi)$ is a cosine.

This paper demonstrates the feasibility of a class of experiments
which can simulate fundamental aspects of models of inflation. Such
experiments, as well as theoretical calculations and numerical simulation,
should follow a step-by-step programme that starts with the simplest
possible theories. In the early stages, naturally, one will learn
a lot about how to extract meaningful data from experiments and the
adequacy of approximate numerical techniques but not much about the
universe. Here we illustrate our approach by a numerical simulation
of quantum effects, using a truncated Wigner approximation described
later. The results show the importance of a hybrid approach, combining
numerical and experimental techniques. Each method involves different
physical approximations. By comparing them, one can hope to come to
an improved understanding of how quantum fluctuations behave in the
nonlinear regime, where exact quantum field predictions are hard to
come by. As the programme is extended to include additional quantum
fields and the effects of gravity one can hope to connect the results
of experiments with astronomical observations and refine our models
of the universe.

In general, the longer term fate of vacuum fluctuations in the nonlinear
regime is of great fundamental interest. It is not known how to calculate
this exactly, due to quantum complexity and the failure of perturbation
theory. We speculate that these nonlinear effects may be related to
very large scale inhomogeneities in the mass distribution of the universe.
This in turn could provide the basis for observational tests of inflationary
universe theories beyond those available now. Unlike some modern models
of inflation~\cite{Linde1991,Linde1994a,Linde1998}, our system can
support domain walls of relative phase~\cite{Son2002}. For experimental
studies, these provide an easily measured signature of vacuum decay,
and may have cosmological implications.

The paper is organized as follows. In Section~(II) we describe the
general theory of coupled Bose fields, as found in Bose-Einstein condensate
(BEC) experiments on ultra-cold atomic gases at nanoKelvin temperatures.
In Section~(III), we treat the different types of classical vacuum
that exist, and the density-phase representation. In Section~(IV)
we transform to the sine-Gordon Lagrangian, and demonstrate how this
can behave as a relativistic field theory with an unstable vacuum.
In Section~(V) we carry out detailed, though approximate, numerical
simulations of the original field theory, to indicate what to expect
in an experiment. Finally, in Section~(VI) we discuss the conclusions.

\section{Coupled Bose fields}

Our first task then, is to slow down light: to a few centimeters per
second if possible. How is this to be achieved? For the purpose, we
use quasi-particles with dispersion relations equivalent to relativistic
equations, in a coupled ultra-cold Bose condensate (BEC). To understand
this, let us consider the dynamical equations for a coupled condensate;
a $D$-dimensional Bose gas with two spin components that are linearly
coupled by an external microwave field. This system obeys the following
non-relativistic Heisenberg equation:

\begin{eqnarray}
i\hbar\partial_{t}\Psi_{1} & = & \left\{ -\frac{\hbar^{2}}{2m}\nabla_{x}^{2}+g_{1}\Psi_{1}^{\dagger}\Psi_{1}+g_{c}\Psi_{2}^{\dagger}\Psi_{2}\right\} \Psi_{1}-\nu\Psi_{2},\nonumber \\
i\hbar\partial_{t}\Psi_{2} & = & \left\{ -\frac{\hbar^{2}}{2m}\nabla_{x}^{2}+g_{2}\Psi_{2}^{\dagger}\Psi_{2}+g_{c}\Psi_{1}^{\dagger}\Psi_{1}\right\} \Psi_{2}-\nu\Psi_{1}.\nonumber \\
 &  & \,
\end{eqnarray}

Here, $\nabla_{x}^{2}$ represents the $D$-dimensional Laplacian
operator, and $g_{1}$, $g_{c}$ are the $D-$dimensional coupling
constants, which are assumed positive. These have known relations
with the measured scattering length. The coupling-constant is renormalizable
at large momentum cutoff, although our simulations are in the regime
of low momentum cutoff, which makes renormalization unnecessary. The
field commutators are:
\begin{equation}
\left[\Psi_{1}\left(\mathbf{x}\right),\Psi_{1}^{\dagger}\left(\mathbf{x}'\right)\right]=\delta^{D}\left(\mathbf{x}-\mathbf{x}'\right)
\end{equation}

We use spinor notation, introducing $\bm{\Psi}=(\Psi_{1},\Psi_{2})^{T}$
and write the energy functional as $W=\int d^{D}x\, w$, where the
energy density $w(\mathbf{x})$ is
\begin{align}
w={} & \frac{\hbar^{2}}{2m}\nabla_{x}\bm{\Psi}^{\dag}\cdot\nabla_{x}\bm{\Psi}-\nu\bm{\Psi}^{\dag}\sigma_{x}\bm{\Psi}+\frac{g_{s}}{2}:(\bm{\Psi}^{\dag}\bm{\Psi})^{2}:+\nonumber \\
 & +\frac{g_{sa}}{2}:(\bm{\Psi}^{\dag}{\sigma_{z}}\bm{\Psi})^{2}:+{g_{a}}:(\bm{\Psi}^{\dag}\bm{\Psi})(\bm{\Psi}^{\dag}{\sigma_{z}}\bm{\Psi}):.
\end{align}
 Here we have defined $2g_{s}=\frac{1}{2}(g_{1}+g_{2})+g_{c}$, $4g_{a}=g_{1}-g_{2}$,
and $2g_{sa}=\frac{1}{2}(g_{1}+g_{2})-g_{c}$. From now on, for simplicity
we will assume the case of symmetric self coupling $g_{1}=g_{2}\equiv g$
, so that $g_{a}=0$. In the case where $g_{c}=g$, we further find
$g_{sa}=0$. It is important for our purposes that the cross-term
$g_{c}$ have a different strength to the self-term $g_{i}$. This
is generally achievable in ultra-cold atomic physics, depending on
the atomic species and particular Feshbach resonance used for magnetically
tunable cases. The fully symmetric case with $g_{c}=g$ is less interesting,
and we will assume that $g_{c}=0$ for definiteness in later numerical
examples.

We recover the Heisenberg equation from $i\partial_{t}\psi_{i}=\delta W/\delta\psi_{i}^{*}$.
Alternatively, we will also recover these equations from a stationary
action principle $\delta S=0$ with
\begin{align}
S=\int dt\, d^{D}x\,[\mathcal{R}\{\bm{\Psi}^{\dag}i\hbar\partial_{t}\bm{\Psi}\}-w] & =\int dt\, d^{D}x\,\left[\mathcal{L}_{B}-w\right].
\end{align}
 The coupling field $\nu$ is supplied by an external microwave source,
and couples the hyperfine levels in the atomic condensate together.
In experimental realizations~\cite{Egorov2011}, this has an adjustable
amplitude and phase, good homogeneity, and a long coherence time.

Before proceeding further, we rescale the equations into natural units.
The time and distance scale is chosen so that mean-field frequency
shifts are of order unity, as are the corresponding Laplacian terms.
For typical atomic densities per Bose field of $n=N/V$, where $N$
is the particle number of a given species, and $V$ the volume, this
leads to the choice $\tau=t/t_{0}$, and $\bm{\zeta}=\mathbf{x}/x_{0}$,
where:
\begin{align}
t_{0} & =\hbar/gn\nonumber \\
x_{0} & =\hbar/\sqrt{gnm}\,.\label{eq:xunit}
\end{align}

Scaling the fields so that they correspond to densities in the new
fields, i.e. $\psi=\Psi x_{0}^{D/2}$, the resulting dimensionless
field equations in the symmetric case are
\begin{align}
i\partial_{\tau}\psi_{1} & =\left\{ -\frac{1}{2}{\nabla}^{2}+\gamma\psi_{1}^{\dagger}\psi_{1}+\gamma_{c}\psi_{2}^{\dagger}\psi_{2}\right\} \psi_{1}-\tilde{\nu}\psi_{2},\nonumber \\
i\partial_{\tau}\psi_{2} & =\left\{ -\frac{1}{2}{\nabla}^{2}+\gamma\psi_{1}^{\dagger}\psi_{2}+\gamma_{c}\psi_{1}^{\dagger}\psi_{1}\right\} \psi_{2}-\tilde{\nu}\psi_{1}.\label{eq:dlGP}
\end{align}
Here, $\nabla$ represents derivatives with respect to $\bm{\zeta}$,
$\gamma=1/(nx_{0}^{D})$, $\gamma_{c}=g_{c}/(gnx_{0}^{D})$, and $\tilde{\nu}=\nu/(gn)$.
The dimensionless field commutators are:
\begin{equation}
\left[\psi_{1}\left(\bm{\zeta}\right),\psi_{1}^{\dagger}\left(\mathbf{\bm{\zeta}}'\right)\right]=\delta^{D}\left(\mathbf{\bm{\zeta}}-\mathbf{\bm{\zeta}}'\right).
\end{equation}

In one dimension, $\gamma^{2}=\gamma_{LL}\equiv mg/(\hbar^{2}n)$,
which is the famous Lieb-Liniger parameter~\cite{Lieb1963} of the
one-dimensional quantum Bose gas.

\section{Stable and unstable vacua}

We now consider a semi-classical approach. We define $\tilde{\psi}_{1}$
as a classical mean field, and we linearize around the classical equilibrium
solutions. Vacuum solutions with constant fields, apart from an oscillating
phase, are easily found from Eq.~(\ref{eq:dlGP}) as $\tilde{\psi}_{1}^{0}=\sqrt{\tilde{n}}=\pm\tilde{\psi}_{2}^{0}$.
An overall phase is chosen to make $\tilde{\psi}_{1}^{0}$ positive
and $\tilde{n}=nx_{0}^{D}$ is the dimensionless density. The different
signs of $\tilde{\psi}_{2}^{0}$ correspond to two inequivalent vacua.
In order to study their stability properties and elementary excitations,
we investigate small oscillations around the vacuum solutions. This
generalizes the analysis presented in Ref.~\cite{Brand2010} to include
the cross-coupling with $\gamma_{c}$.

\subsection{Linearized solutions}

Writing $\tilde{\psi}_{j}=[\tilde{\psi}_{j}^{0}+(u_{k}^{j}e^{i(\tilde{\mathbf{k}}\bm{\zeta}-\tilde{\omega}\tau)}+v_{k}^{j\ast}e^{-i(\tilde{\mathbf{k}}\bm{\zeta}-\tilde{\omega}\tau)})]e^{-i\tilde{{\color{red}{\normalcolor \mu}}}\tau}$,
where $\tilde{\mu}$ is the vacuum chemical potential, substituting
into Eq.~(\ref{eq:dlGP}) and keeping linear terms we obtain the
secular equation $\det[B-\tilde{\omega}]=0$ with
\begin{align}
B=\left(\begin{array}{cccc}
H_{0} & \gamma\tilde{n} & \mp\tilde{\nu}+\gamma_{c}\tilde{n} & \gamma_{c}\tilde{n}\\
-\gamma\tilde{n} & -H_{0} & -\gamma_{c}\tilde{n} & \pm\tilde{\nu}-\gamma_{c}\tilde{n}\\
\mp\tilde{\nu}+\gamma_{c}\tilde{n} & \gamma_{c}\tilde{n} & H_{0} & \gamma\tilde{n}\\
-\gamma_{c}\tilde{n} & \pm\tilde{\nu}-\gamma_{c}\tilde{n} & -\gamma\tilde{n} & -H_{0}
\end{array}\right),
\end{align}
 where $H_{0}=-\frac{1}{2}\tilde{k}^{2}+(2\gamma+\gamma_{c})\tilde{n}-\tilde{\mu}$
and consistency demands that $2\gamma_{s}\tilde{n}=\tilde{\mu}\pm\tilde{\nu}$,
where we define $2\gamma_{s}=\gamma+\gamma_{c}$. The sign \textquotedbl{}+\textquotedbl{}
corresponds to the in-phase and \textquotedbl{}-\textquotedbl{} corresponds
to the out-of-phase vacua. We find two independent solution branches
of the secular equation. The first one is gapless
\begin{equation}
\tilde{\omega}_{1}=\sqrt{\frac{1}{2}\tilde{k}^{2}\left(\frac{1}{2}\tilde{k}^{2}+4\gamma_{s}\tilde{n}\right)},
\end{equation}
with eigenvector $\sim(1,-1,1,-1)^{{\rm T}}$ for small $k$, which
corresponds to the phase fluctuations of the fields $\delta\arg\tilde{\psi}_{1}=\delta\arg\tilde{\psi}_{2}\propto\sin(\tilde{\mathbf{k}}\bm{\zeta}-\tilde{\omega_{1}}\tau)$.
This leads to a sound wave with sound speed $\tilde{v}_{\mathrm{Bog}}=\sqrt{2\gamma_{s}\tilde{n}}$.
Solutions with real frequencies indicate stable, propagating waves
of small amplitude, whereas imaginary roots indicate an instability.
Excitations along this branch are always stable when $2\gamma_{s}\equiv\gamma+\gamma_{c}>0$.

The other branch is gapped, since
\begin{equation}
\tilde{\omega}_{2}=\sqrt{\left(\frac{1}{2}\tilde{k}^{2}\pm2\tilde{\nu}\right)\left(\frac{1}{2}\tilde{k}^{2}+4\gamma_{sa}\tilde{n}+2\tilde{\nu}\right)},\label{eq:omega2}
\end{equation}
 where $2\gamma_{sa}=\gamma-\gamma_{c}$. The nonlinearity parameter
in this second branch differs from the first branch and can be tuned
independently due to the presence of the cross-coupling $\gamma_{c}$.
The eigenvector corresponding to $\omega_{2}$ is $\sim(1,-1,-1,1)^{{\rm T}}$
for small $k$ and $\nu$, which corresponds to an excitation of relative
phase fluctuations of the fields $\delta\arg\tilde{\psi}_{1}=-\delta\arg\tilde{\psi}_{2}\propto\sin(\tilde{\mathbf{k}}\bm{\zeta}-\tilde{\omega}_{2}\tau)$.
It is interesting to note that the two branches are decoupled in the
linear approximation. We will be particularly concerned with the relative
phase dynamics, which has characteristic properties of the elementary
excitations of a relativistic field theory. Indeed, expanding to second
order in $\tilde{k}$ we may write $\tilde{\omega}_{2}^{2}=\tilde{k}^{2}\tilde{c}_{2}^{2}+\tilde{m}_{2}^{2}\tilde{c}_{2}^{4}$,
which is a relativistic dispersion relation with a light speed of
$\tilde{c}_{2}^{2}=2\gamma_{sa}\tilde{n}\pm2\tilde{\nu}$ and rest
energy $\tilde{m}_{2}^{2}\tilde{c}_{2}^{4}=4\tilde{\nu}(2\gamma_{sa}\tilde{n}\pm\tilde{\nu})$.
The relative phase dynamics will analyzed be more rigorously in the
next section.

The second branch also supports unstable modes and governs the dynamics
of the unstable vacuum. We are particularly interested in the situation
where both $\gamma_{s}$ and $\gamma_{sa}$ are positive. In this
case the second branch becomes unstable if the linear coupling is
such that $\pm\tilde{\nu}<0$, with the unstable modes satisfying
$-4\gamma_{sa}\tilde{n}\mp2\tilde{\nu}<\tilde{k}^{2}/2<\mp2\tilde{\nu}$.
This occurs either with the fields having the same sign and $\tilde{\nu}<0$,
or with the fields having the opposite sign and $\tilde{\nu}>0$.

Thus, in a sufficiently large system where $k$ is continuous, one
of the two possible vacua is stable and the other one unstable. The
classical field dynamics resulting from the unstable vacuum has been
discussed for some special cases in Refs.~\cite{Lesanovsky2007,Montgomery2010}
(see also Ref.~\cite{Brand2010}). In this paper we are going to
simulate the decay from the unstable vacuum due to quantum fluctuations.
The characteristic length scale of the decay modes is larger than
$\pi/\sqrt{\left|\tilde{\nu}\right|}$, which will become the largest
length scale in the system in the interesting regime where $|\tilde{\nu}|$
is small. The time scales of the unstable mode become large in the
sine-Gordon regime. This is compatible with the following inflation
requirement: When the scalar field rolls down the potential hill very
slowly compared to the expansion of the universe, inflation occurs.

\subsection{Density-phase representation}

We will continue to consider these equations as classical, or mean-field
equations. We use this procedure to give us a better understanding
of how the dynamics can be reduced to those of a relativistic scalar
field. Once we have made this transformation, we will use Lagrangian
quantization methods to construct a low-energy effective quantum field
theory for the elementary excitations.

\begin{figure}
\includegraphics[width=0.9\columnwidth]{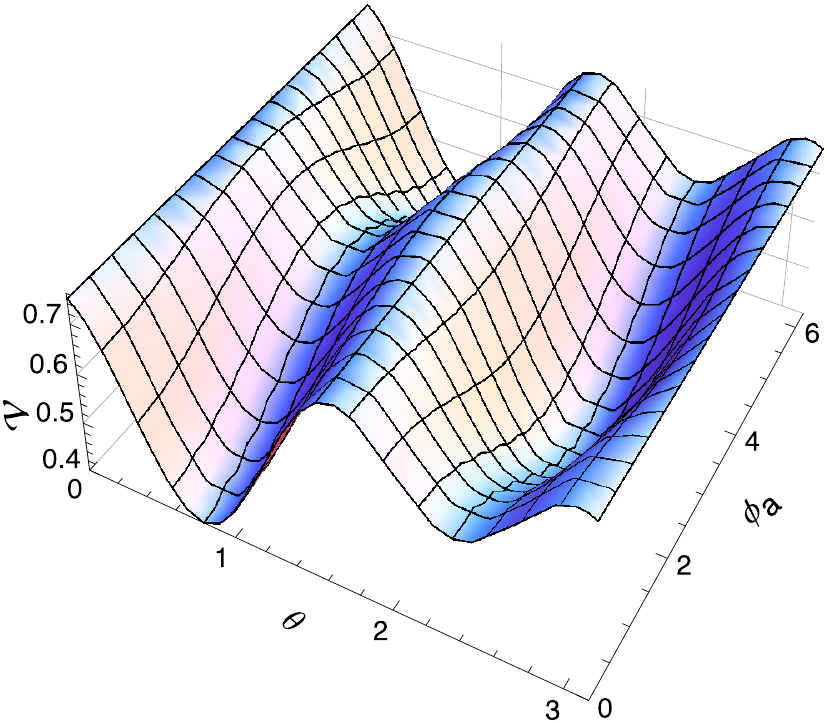} \caption{\label{fig:cut1}Plotted is $\mathcal{V}$ for $\tilde{\nu}=0.1$,
$u=1$, $\gamma_{sa}=0.5$ as a function of $\theta$ and $\phi_{a}$.
At $\theta=3\pi/4$ a saddle point occurs, which corresponds to $\pi$
phase jump between the two components. It is a known and dynamically
unstable stationary state of the GP equations~\cite{Lesanovsky2007,Brand2010}.}
\end{figure}

We parametrize the spinor as follows:
\begin{align}
\tilde{\psi}_{1} & =ue^{i(\phi_{s}+\phi_{a})/2}\cos(\theta)\nonumber \\
\tilde{\psi}_{2} & =ue^{i(\phi_{s}-\phi_{a})/2}\sin(\theta),\label{eq:dpdefn}
\end{align}
 by introducing the density mixing angle $\theta$ and total density
$u^{2}$, the relative phase $\phi_{a}$ and total phase $\phi_{s}$.
The first part of the dimensionless Lagrange density becomes
\begin{align}
\tilde{\mathcal{L}}_{B} & =\frac{t_{0}x_{0}}{\hbar}\mathcal{L}_{B}=\mathcal{R}\{\tilde{\bm{\psi}}^{\dag}i\partial_{\tau}\tilde{\bm{\psi}}\}\nonumber \\
 & =\frac{u^{2}}{2}\partial_{\tau}\phi_{s}+\frac{u^{2}}{2}\cos(2\theta)\partial_{\tau}\phi_{a}.
\end{align}
 Writing the energy density as $\tilde{w}=\tilde{\mathcal{K}}+\tilde{\mathcal{V}}$,
the kinetic energy part becomes
\begin{align}
\tilde{\mathcal{K}}= & \frac{1}{2}{\nabla}\tilde{\bm{\psi}}^{\dag}\cdot{\nabla}\tilde{\bm{\psi}}\nonumber \\
= & \frac{1}{2}\Big\{({\nabla}u)^{2}+u^{2}({\nabla}\theta)^{2}+\left[({\nabla}\phi_{s})^{2}+({\nabla}\phi_{a})^{2}\right]\frac{u^{2}}{4}\nonumber \\
 & +{\nabla}\phi_{s}\cdot{\nabla}\phi_{a}\frac{u^{2}}{4}\cos(2\theta)\Big\}.
\end{align}
 For the potential part we find
\begin{align}
\mathcal{V}= & -\tilde{\nu}u^{2}\cos\phi_{a}\sin(2\theta)+\frac{\gamma_{s}}{2}u^{4}\nonumber \\
 & +\frac{\gamma_{sa}}{4}u^{4}[1+\cos(4\theta)],
\end{align}
 where $2\gamma_{s}=\gamma+\gamma_{c}$ and $2\gamma_{sa}=\gamma-\gamma_{c}$.
The potential is a function of the three real field variables $u$,
$\theta$, and $\phi_{a}$. A cut for constant $u$ is shown in Fig.~\ref{fig:cut1}.
Note:
\begin{itemize}
\item As long as $\gamma_{sa}\ge0$, there is a valley at $\theta=\pi/4$
corresponding to a vacuum with equal density in the two fields. This
expresses a well-known condition for the stability of two coupled
Bose fields, $g_{1}+g_{2}\ge2g_{c}$, where only the total particle
number is conserved. This condition is weaker than the miscibility
condition for binary mixtures without interconversion~\cite{Esry1997a},
$g_{1}g_{2}>g_{c}^{2}$, and turns into it when $g_{1}=g_{2}$. The
vacuum is located at $(\theta=\pi/4,\phi_{a}=0)$. An equivalent point
is $(\theta=3\pi/4,\phi_{a}=\pi)$. Working at fixed particle number
$N$, we find for the dimensionless total density of the vacuum $u^{2}=2\tilde{n}\equiv2nx_{0}^{D}=\gamma^{-1}$.
\item As $\theta$ terms primarily couple to the relative phase $\phi_{a}$,
there are dynamical solutions following the bottom of the $\theta$
valley with $\phi_{a}=0$ and $\theta=\pi/4$ corresponding to solutions
of the type $\psi_{1}(\mathbf{x},t)=\psi_{2}(\mathbf{x},t)$. This
is just the dynamics of a single $D$-dimensional Bose field involving
compressional waves (Bogoliubov phonons) and dark solitons.
\item The point $(\theta=\pi/4,\phi_{a}=\pi)$ or, equivalently, $(\theta=3\pi/4,\phi_{a}=0)$
is an unstable point.
\item Excitations of the relative phase $\phi_{a}$ have low energy cost
when $|\tilde{\nu}|\ll\gamma_{sa}u^{2}$, i.e.\ in the regime of
small coupling and strong nonlinearity. This is the main focus of
this paper.
\end{itemize}

We can see from Fig.~\ref{fig:cut1} that within each large canyon
in $\theta$-space, there is a secondary feature where a hill and
valley form in the $\phi-$direction. We wish to focus on this relative
phase feature. It is able to form a quantum field with a relativistic
dispersion relation, albeit with an effective ``light'' velocity
of very much smaller size than $c$.

\subsection{Domain walls}

Domain walls of relative phase appear as a connection between two
degenerate vacua along a $\theta$ valley. The corresponding solution
were discussed in the context of three-dimensional two-component BECs
with internal coupling in Ref.~\cite{Son2002}. In the context of
one dimensional coupled BECs they are also known as atomic Josephson
vortices or rotational fluxons, since they are related to circular
atomic currents, and were discussed in Refs.~\cite{Kaurov2005,Kaurov2006,Brand2009,Qadir2012}.
As features of the relative phase, they only exist at sufficiently
small linear coupling $\tilde{\nu}<1/3$~\cite{Kaurov2005}. They
are unstable and can decay if a mechanism for dissipation of energy
is provided unless $\tilde{\nu}\lesssim0.14$, where they become local
energy minima with topological stability~\cite{Su2013}. In this
regime, they acquire the typical properties of topological solitons,
or kinks, in the sine-Gordon equation. By tuning the coupling parameter
$\tilde{\nu}$, one can thus move between regimes of different stability
properties of domain walls. The regime dominated by sine-Gordon physics
is found for small $\tilde{\nu}$.

\section{Sine-Gordon regime}

Aiming at the dynamics of the relative phase $\phi_{a}$, we simplify
the action and corresponding equations of motion by assuming that
all dynamics takes place in a $\theta$ valley and that both $\theta$
and $u$ only have small deviations from their respective equilibrium
values. We consequently write $\theta=\pi/4+y$ and expand the action
to leading orders in $y$. Using $\cos(2\theta)=-2y$, $\sin(2\theta)=1-2y^{2}$,
$\cos(4\theta)=8y^{2}-1$ we obtain
\begin{subequations}
\begin{align}
\tilde{\mathcal{V}}={} & \tilde{\nu}u^{2}\cos\phi_{a}(1-2y^{2})-\frac{\gamma_{s}}{2}u^{4}-2{\gamma_{sa}}u^{4}y^{2}\nonumber \\
2\tilde{\mathcal{K}}={} & ({\nabla}u)^{2}+u^{2}({\nabla}y)^{2}+\left\{ ({\nabla}\phi_{s})^{2}+({\nabla}\phi_{a})^{2}\right\} \frac{u^{2}}{4}\nonumber \\
 & -{\nabla}\phi_{s}{\nabla}\phi_{a}\frac{u^{2}}{2}y\nonumber \\
\tilde{\mathcal{L}}_{B}={} & \frac{u^{2}}{2}\partial_{\tau}\phi_{s}-{u^{2}}y\partial_{\tau}\phi_{a}.
\end{align}

\end{subequations}
Low energy dynamics is possible in two sectors: compressional waves
involving $u$ and $\phi_{s}$ on the one hand, and dynamics of the
relative phase involving $\phi_{a}$ and $y$ on the other. Keeping
only the terms that are relevant for the latter, the Lagrangian density
reads
\begin{align}
\tilde{\mathcal{L}}=-u^{2}(\partial_{\tau}\phi_{a})y-\frac{u^{2}}{8}({\nabla}\phi_{a})^{2}+\tilde{\nu}u^{2}\cos(\phi_{a})-2\gamma_{sa}u^{4}y^{2}.
\end{align}
 Variation with respect to the fields gives the Euler-Lagrange equation
\begin{align}
\frac{\partial\mathcal{L}}{\partial f}-{\nabla}\frac{\partial\mathcal{L}}{\partial{\nabla}f}-\partial_{\tau}\frac{\partial\mathcal{L}}{\partial f_{\tau}}=0,
\end{align}
 which implies that$\partial_{\tau}\phi_{a}+4\gamma_{sa}u^{2}y=0$.
This allows us to replace $y$ in the Lagrangian to yield
\begin{align}
\tilde{\mathcal{L}}=\frac{1}{8\gamma_{sa}}(\partial_{\tau}\phi_{a})^{2}-\frac{\tilde{n}}{4}({\nabla}\phi_{a})^{2}+2\tilde{\nu}\tilde{n}\cos(\phi_{a}).\label{eq:sine-Gordon-Lagrangian}
\end{align}
 This is the {$D$-dimensional sine}-Gordon Lagrangian.

Now we need to extract parameters. The critical speed $\tilde{c}$
is found by dividing the ${\nabla}$ pre-factor by the $\partial_{\tau}$
pre-factor and taking the square root to yield $\tilde{c}=\sqrt{2\gamma_{sa}\tilde{n}}$.
The critical velocity thus matches the Bogoliubov speed of sound $\tilde{v}_{\mathrm{bog}}=\sqrt{2\gamma_{s}\tilde{n}}$
only in the case of vanishing cross coupling $\gamma_{c}$, where
$\gamma_{sa}=\gamma_{s}$. We note that so far we have used classical
arguments. However, our original equations are also valid as quantum
field equations. If we quantize the Lagrangian, we accordingly have
a quantum sine-Gordon equation, which describes the physics at sufficiently
low real temperatures.

On quantizing the Lagrangian, Eq.~(\ref{eq:sine-Gordon-Lagrangian}),
the canonical momentum field is:
\[
\pi=\frac{\partial_{\tau}\phi_{a}}{4\gamma_{sa}}\,,
\]
 with corresponding canonical commutators of:
\[
\left[\phi_{a}\left(\bm{\mathbf{\zeta}}\right),\pi\left(\bm{\mathbf{\zeta}}'\right)\right]=i\delta^{D}\left(\bm{\mathbf{\zeta}}-\bm{\mathbf{\zeta}}'\right)\,.
\]

The sine-Gordon action can be brought into Lorenz-invariant form by
rescaling time as $\zeta_{0}=\tilde{c}\tau$ to read
\begin{align}
S=\hbar\int d^{D+1}\zeta\,\frac{1}{2}\left\{ (\partial_{\zeta_{0}}\phi')^{2}-({\nabla}\phi')^{2}\right\} +\frac{\tilde{\alpha}}{\beta^{2}}\cos(\beta\phi'),
\end{align}
 with the rescaled field $\phi_{a}=\beta\phi'$ and $\beta^{2}=2\sqrt{2\gamma_{sa}\gamma},\,\,\tilde{\alpha}=4\tilde{\nu}$.
Here, $\beta^{2}$ is the universal dimensionless parameter of the
quantum sine-Gordon equation. In one dimension and the case of vanishing
cross coupling $\gamma_{c}=0$ and $2\gamma_{sa}=\gamma$, we obtain
$\beta^{2}=2\gamma=2\sqrt{\gamma_{LL}}$, which again links to the
Lieb-Liniger coupling parameter $\gamma_{LL}=mg/(\hbar^{2}n)$. 

The stationary action principle leads to the (classical) sine-Gordon
equation
\begin{align}
{{\nabla}^{2}\phi_{a}-\partial_{\zeta_{0}\zeta_{0}}\phi_{a}+\tilde{\alpha}\sin\phi_{a}=0.}
\end{align}
 Note that the original field $\phi_{a}$ reappears and the parameter
$\beta$ completely drops out of the classical equation. This is a
relativistic field equation with a dimensionless speed of light of
unity, as in all sensible relativistic field theories. Alternatively,
using atomic units, we consider $\tilde{c}$ as the effective light-speed.

The parameter $\alpha=\tilde{\alpha}x_{0}^{-2}$ is connected to the
characteristic length scale of the sine-Gordon equation. In physical
units
\begin{align}
\ell_{SG}=\frac{1}{\sqrt{{\alpha}}}=\frac{\hbar}{2\sqrt{{\nu}m}}.
\end{align}
 Remarkably, this length scale depends on the tunnel coupling but
is independent of the nonlinearity. It is the second characteristic
length scale besides the GP healing length $x_{0}$ of Eq.~(\ref{eq:xunit}).
It sets the spatial scales of the domain walls in the sine-Gordon
regime.

The parameter $\tilde{\alpha}$ is also connected to the rest mass
of the elementary excitations of the sine-Gordon equation. Indeed,
small amplitude oscillations $\sim\exp{-i(\tilde{\mathbf{k}}\bm{\zeta}-\tilde{\omega}_{sG}\tau)}$
of the vacuum have a relativistic dispersion relation of $\tilde{\omega}_{sG}^{2}=\tilde{k}^{2}\tilde{c}^{2}+\tilde{\alpha}\tilde{c}^{2}$,
with a rest mass of $\tilde{m}_{sG}=\sqrt{\tilde{\alpha}}/\tilde{c}$
and a rest energy of $\tilde{m}_{sG}^{2}\tilde{c}^{4}=8\tilde{\nu}\gamma_{sa}\tilde{n}$.
Note that the sine-Gordon light speed $\tilde{c}$ and rest mass $\tilde{m}_{sG}$
agree to leading order in $\tilde{\nu}$ with the corresponding parameters
of the gapped second branch of the Bogoliubov spectrum of the stable
vacuum~(\ref{eq:omega2}). The relativistic sine-Gordon equation
thus describes asymptotically a sector of excitations of the vacuum
that, for small amplitude, completely decouple from the compressional
waves of the Bogoliubov sound.

In the course of inflation tiny quantum fluctuations are important:
they form the primordial seeds for all structure created in the later
Universe. We now turn to computational quantum simulations of these
equations.

\section{Computational Simulations}

To investigate the feasibility and likely behavior of these experiments,
we have carried out a number of numerical simulations of the expected
dynamics. We simulated the full coupled Bose fields, using atomic
densities, confinement parameters and S-wave scattering lengths generally
similar to those employed in recent experimental studies with ultra-cold
atoms. We employed an accurate fourth-order interaction-picture algorithm.
Careful monitoring of step-size errors was needed to ensure accurate
results.

The theoretical method used is a truncated, probabilistic version
of the Wigner-Moyal~\cite{Wigner1932,Moyal1947} phase-space representation~\cite{Drummond1993,Steel1998,Sinatra2002,Blakie2008,Opanchuk2013}.
This is a probabilistic phase-space method for quantum fields, that
correctly simulates symmetrically-ordered quantum dynamics in the
limit of large numbers of atoms per mode. It has been shown to give
results in agreement with quantum-limited photonic and BEC experiments~\cite{Corney2008,Opanchuk2012},
provided the large atom-number restriction is satisfied in the experiment.

Other methods of this type are also possible. One example is the positive-P
representation~\cite{Drummond1980,Carter1987}. In this normal-ordered
approach, there is no truncation. Unlike the truncated Wigner method,
no systematic errors occur at small occupation numbers~\cite{Deuar2007}.
However, with current algorithms, the statistical sampling error grows
in time for long time-scale nonlinear problems without damping. This
rules out the positive-P approach for the present simulations, until
better algorithms are found.

Although the resulting equations are similar to the Gross-Pitaevskii
(GP) equations, the initial noise terms have a precise meaning from
quantum mechanics. They correspond to the exact vacuum noise needed
to generate symmetrically ordered operator moments that include the
quantum fluctuations of the initial quantum state. The truncated Wigner
method therefore includes quantum terms correct to order $1/n$, which
are omitted in the GP approach. Additional stochastic terms are needed
if there is decoherence or atomic absorption~\cite{Opanchuk2013}.
However, in these preliminary studies, such effects will be omitted.
We also omit the cross-coupling between spins for simplicity. Following
the dimensional notation introduced previously, except that our c-number
fields are now stochastic representations of quantum fields, we obtain:
\begin{align}
i\partial_{\tau}\tilde{\psi}_{1} & =\left\{ -\frac{1}{2}{\nabla_{\zeta}}^{2}+\gamma\left|\tilde{\psi}_{1}\right|^{2}\right\} \tilde{\psi}_{1}-\tilde{\nu}\tilde{\psi}_{2},\nonumber \\
i\partial_{\tau}\tilde{\psi}_{2} & =\left\{ -\frac{1}{2}{\nabla_{\zeta}}^{2}+\gamma\left|\tilde{\psi}_{2}\right|^{2}\right\} \tilde{\psi}_{2}-\tilde{\nu}\tilde{\psi}_{1}.
\end{align}

For the truncated Wigner calculations, with an initial coherent state
of $\bar{\psi}(\zeta)$, so that $\tilde{\psi}(\zeta)=\bar{\psi}(\zeta)+\Delta\tilde{\psi}(\zeta)$,
one would have:
\[
\left\langle \Delta\tilde{\psi}(\zeta)\Delta\tilde{\psi}^{*}(\zeta')\right\rangle =\frac{1}{2}\delta\left(\zeta-\zeta'\right)\,.
\]
 We emphasize here that our assumption of an initial coherent state
is certainly not the only one possible. It can be replaced by other
quantum states. For example, a finite temperature ground-state followed
by a sudden jump in phase is probably closer to the way that experiments
will take place. Nevertheless, in the absence of a good model for
the quantum state of the early universe, a coherent state is as reasonable
a choice as any other.

Experimental state preparation is a complex issue, and depends on
such details as the magnitude of the microwave coupling during cooling.
Our initial state corresponds approximately to one in which the fields
are cooled to a very weakly interacting ground state, with a coherent
coupling of $\nu<0$, so that the lowest energy state has fields with
an opposite sign. Then the interactions are increased to a large value,
and the coupling phase is reversed to $\nu>0$, in order to create
an unstable vacuum. Naturally, other models are possible, including
a strongly interacting initial state at finite temperature. These
will be treated elsewhere.

\subsection{Dependence on couplings}

\begin{figure}
\includegraphics[width=1\columnwidth]{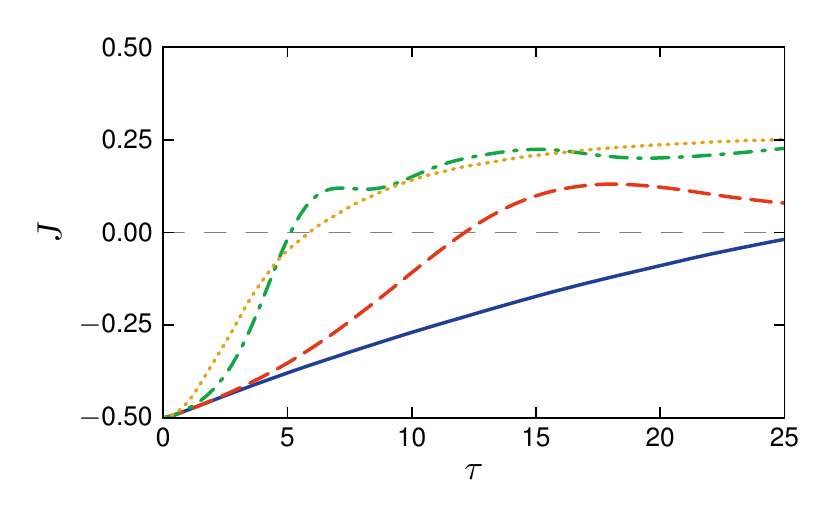}

\caption{Decay of 1D quantum field coherence $J$, as an indicator of Sine-Gordon
unstable dynamics. $\gamma=0.1$. Couplings, $\tilde{\nu}=0.001,\,0.01,\,0.1,\,1$
(blue solid, red dashed, green dash-dotted, yellow dotted). Condensate
size $L=80$, 256 spatial grid points, 10000 time steps, 100 ensembles,
opposite initial phases to give an unstable vacuum.\label{fig:Decay-of-1DvsCouplings}}
\end{figure}

We firstly investigate the dependence of the system on the coupling
strength. This is a crucial part of the calculation, as it creates
the potential minimum such that the relative phase is able to be treated
as an independent scalar field. The measurable quantities in an experiment
are the densities obtained after interfering the two hyperfine field
amplitudes~\cite{Egorov2011}. These are the quantities:
\begin{equation}
n_{\pm}=\frac{1}{2}\left\langle \left(\hat{\psi}_{1}\pm\hat{\psi}_{2}\right)^{\dagger}\left(\hat{\psi}_{1}\pm\hat{\psi}_{2}\right)\right\rangle \label{eq:npn}
\end{equation}

They correspond simply to the atomic densities after a rapid Rabi
rotation of $\pi/2$ is performed to allow the atomic fields to interfere.
Substituting the mean fields of Eq.~(\ref{eq:dpdefn}) into Eq.~(\ref{eq:npn}),
these measurable quantities are
\[
n_{\pm}=u^{2}\left[\frac{1}{2}\pm\frac{1}{2}\cos\phi_{a}+O\left(y^{2}\right)\right].
\]
At the vacuum phase, $\phi_{a}=0$, the atoms are all in the even
mode, with $n_{+}=u^{2}$ and $n_{-}=0$. At the unstable equilibrium,
$\phi_{a}=\pi$, $n_{+}=0$, and $n_{-}=u^{2}$. We plot the relative
visibility, which is a quadrature-like measure of the phase, namely~\cite{He2012a}:
\[
J=\frac{\int d^{D}\bm{\zeta}\left(n_{+}-n_{-}\right)}{4N}\approx\frac{1}{2}\overline{\cos\phi_{a}}\,.
\]

We see in Fig.~\ref{fig:Decay-of-1DvsCouplings}, that for very weak
couplings, there is a relative phase decay induced purely by quantum
phase diffusion. This is the opposite to sine-Gordon physics, and
occurs because the two condensates are uncoupled, and experience an
increasingly random relative phase. For stronger couplings, with $\tilde{\nu}\approx0.1$,
one can see the distinctive rapid decay of an unstable vacuum, which
is the signature of sine-Gordon physics with this initial condition.
The slight oscillation near $\tilde{\nu}\approx0.1$ may indicate
some form of collective instability of coupled domains, and deserves
further investigation.

\subsection{Effects of condensate size}

Although experimentally one would use a finite trap, we simulate a
system with periodic boundary conditions for simplicity. To some extent,
this allows us to ignore the detrimental effects of boundaries, and
hence to allow a small computer simulation mimic a larger experimental
condensate. However, this is not entirely the case.

\begin{figure}
\includegraphics[width=1\columnwidth]{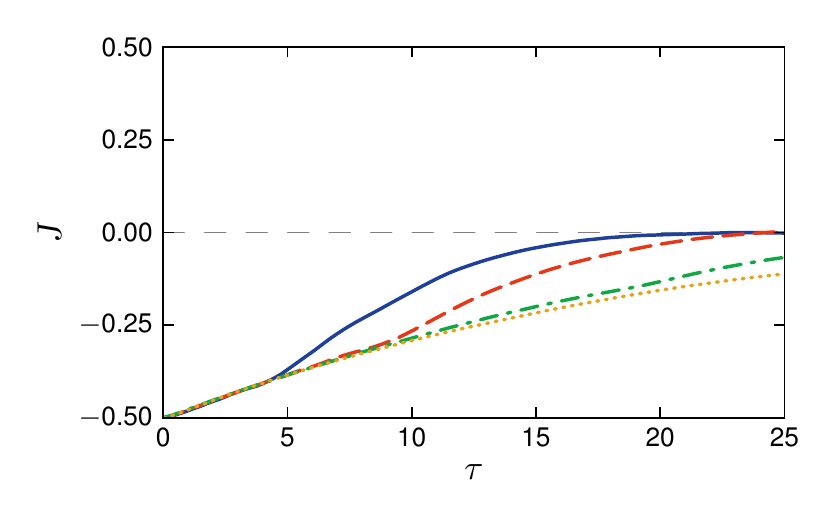}

\caption{Decay of average 1D quantum field coherence $J$ for different condensate
sizes, to indicate the effects of boundaries. Condensate size $L=10,\,20,\,40,\,80$
(blue solid, red dashed, green dash-dotted, yellow dotted), spatial
grid sizes $32,\,64,\,128,\,256$, 10000 time steps, 100 ensembles,
$\gamma=0.1$. Coupling $\tilde{\nu}=0$. Opposite initial phases.\label{fig:Plot-of-1D vs Condensate size}}
\end{figure}

As shown in Fig.~\ref{fig:Plot-of-1D vs Condensate size}, the finite
coherence length of the quantum fluctuations has an effect on the
decay statistics. Initially, the condensate size has no effect, as
the coherence length is very small, and certainly much smaller than
the condensate size as long as it is larger than a healing length.
For increasing times, if $\tilde{\nu}=0$, the coherence length increases
until it reaches the size of the condensate itself. At this stage,
the decay accelerates. For very large condensates, a boundary-independent
behavior is found. We note that this effect is strongest in the limit
of zero coupling $\tilde{\nu}$, which is plotted here. At larger
couplings the formation of domain walls limits the growth of coherence.

\subsection{Single trajectory examples}

Quantum mechanics only predicts ensemble averages. However, it is
now common to use quantum theory to predict the behavior of the universe,
which is only a single ensemble member. Of course, this experiment
is not easy to repeat, especially as the observer is part of it. In
the laboratory, one can obtain ensemble averages by repeating the
state preparation. Even so, the density patterns and evolution in
a single ensemble member is instructive. We expect the Wigner ensemble
members to have a characteristic behavior like individual laboratory
runs, and perhaps to the universe itself, if current theories are
correct.

\begin{figure}
\includegraphics[width=1\columnwidth]{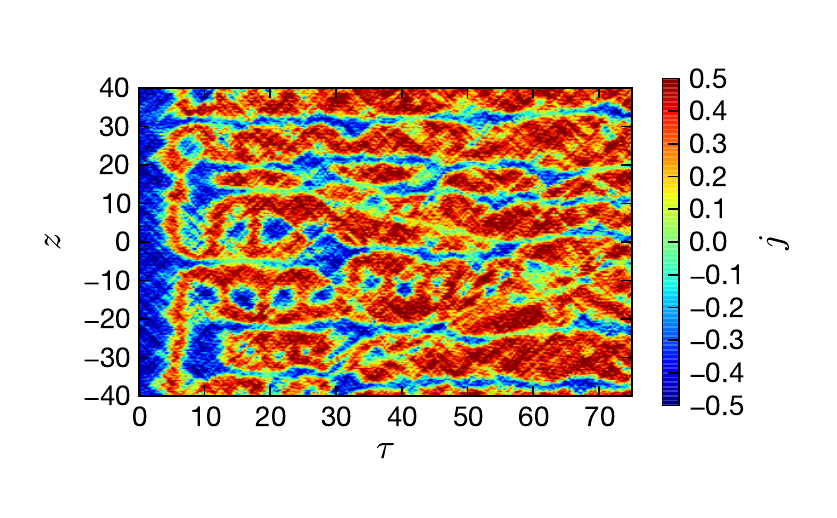}

\caption{Plot of 1D space-time dynamics of the local visibility $j$ for a
single trajectory, showing contour plot vs time and space. Coupling
$\tilde{\nu}=0.1$, condensate size 80, 256 spatial grid points, $\gamma=0.1$,
10000 time steps.\label{fig:Plot-of-1D singletrajectory}}
\end{figure}

In the one-dimensional case, this is shown in Fig.~\ref{fig:Plot-of-1D singletrajectory}.
This corresponds to the largest of the condensates plotted in Fig.~\ref{fig:Plot-of-1D vs Condensate size},
and it is visually clear that the density correlation length is still
substantially smaller than the condensate size throughout the time
evolution. The local fringe visibility is defined in general as:
\[
j=\frac{n_{+}-n_{-}}{2\left(n_{+}+n_{-}\right)}\approx\frac{1}{2}\cos\phi_{a}\,.
\]

Since the total density is nearly constant throughout, we used the
excellent approximation of $j\approx\left(n_{+}-n_{-}\right)/(2\bar{u}^{2})$
for simplicity of calculation and plotting.

\subsection{Two-dimensional behavior}

The universe is certainly not one-dimensional. Anyone might reasonably
question the applicability of a one-dimensional simulation, even though
this is the most amenable to theoretical treatment. In the laboratory,
the availability of engineered traps means that dimensionality can
be readily adjusted to obtain artificial universes of one, two or
three space dimensions.

\begin{figure}
\includegraphics[width=1\columnwidth]{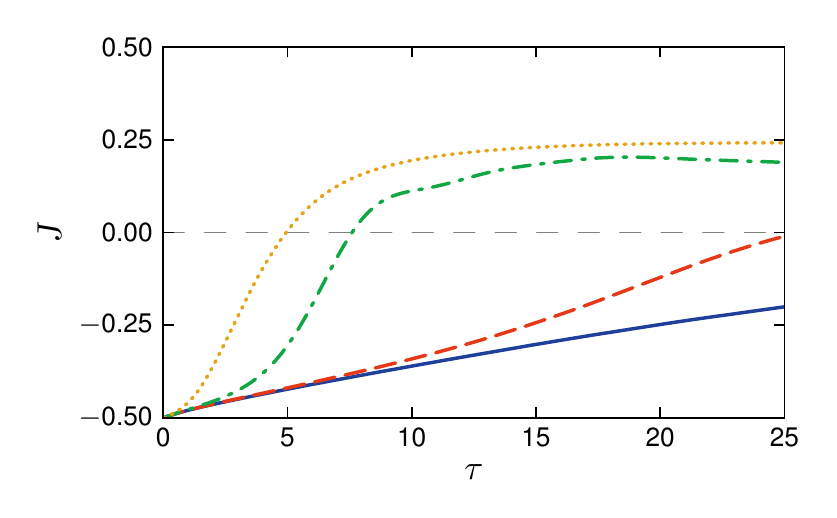}

\caption{Plot of 2D quantum field coherence $J$, as an indicator of sine-Gordon
unstable dynamics. $\gamma=0.1$. Couplings, $\tilde{\nu}=0.001,\,0.01,\,0.1,\,1$
(blue solid, red dashed, green dash-dotted, yellow dotted). Condensate
size $L=80$, $256\times256$ spatial grid points, 4000 time steps,
200 ensembles, opposite initial phases.\label{fig:Plot-of-2D coherence}}
\end{figure}

\begin{figure*}
\includegraphics[width=1\textwidth]{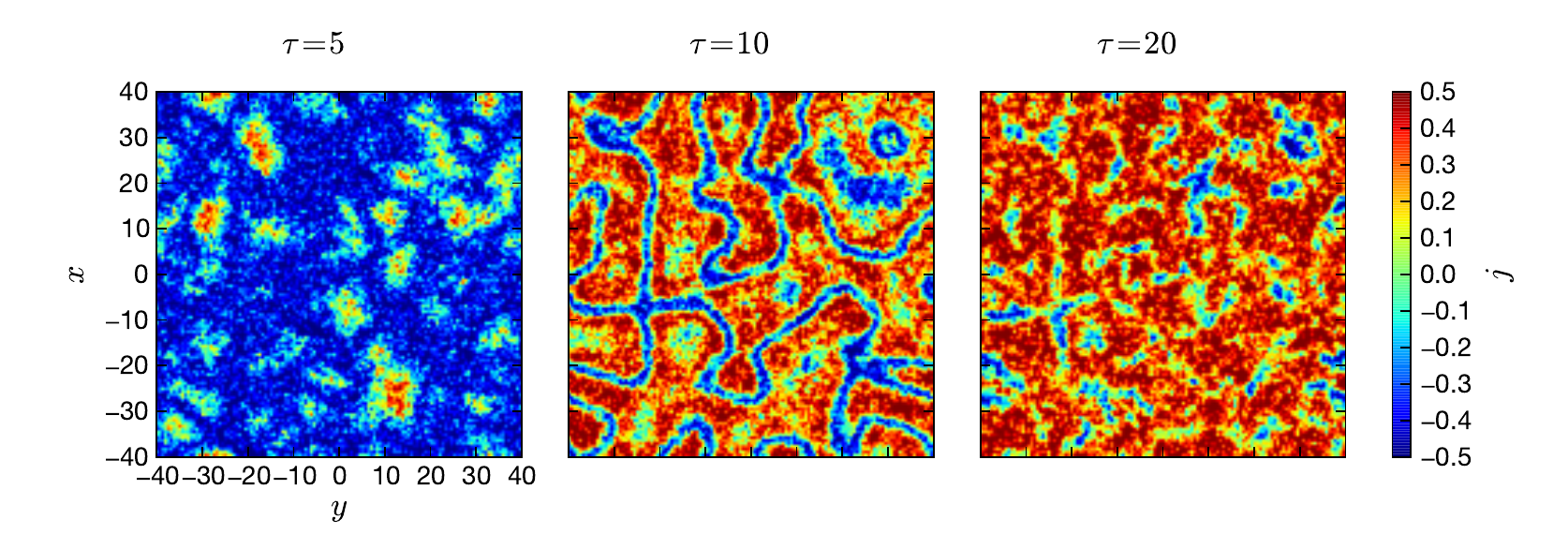}

\caption{\label{fig:Plot-of-2D singletrajectory}Plot of 2D spacial clustering
of the local visibility $j$ for a single trajectory, showing contour
plot vs space at final time. Coupling $\tilde{\nu}=0.1$, condensate
size $80\times80$, $128\times128$ spatial grid points, $\gamma=0.1$,
5000 time steps, opposite initial phases, time $\tau=10$. }
\end{figure*}

The average dynamics of the sine-Gordon field depend on dimensionality
as shown in Fig.~\ref{fig:Plot-of-2D coherence}, which shows that
the two-dimensional averages differ from Fig.~\ref{fig:Decay-of-1DvsCouplings},
although there are qualitative similarities. However, Fig.~\ref{fig:Plot-of-2D singletrajectory}
is much more interesting, demonstrating clearly the growth of complex
spatial structures. These are two-dimensional cross-sections, taken
at dimensionless times of $\tau=5$,$\tau=10$ and$\tau=20$. These
single-trajectory plots show how the early, relatively high momentum
quantum noise patterns have decayed into coherent spatial domains
with a typical spatial coherence length of order $\bar{\zeta}=10$
spatial units at $\tau=10$, and start to evaporate at later times.
It remains to be investigated whether this later evaporation is an
artifact of the coupled BEC system, or a genuine sine-Gordon feature.

\subsection{Three-dimensional behavior}

\begin{figure*}
\includegraphics[width=1\textwidth]{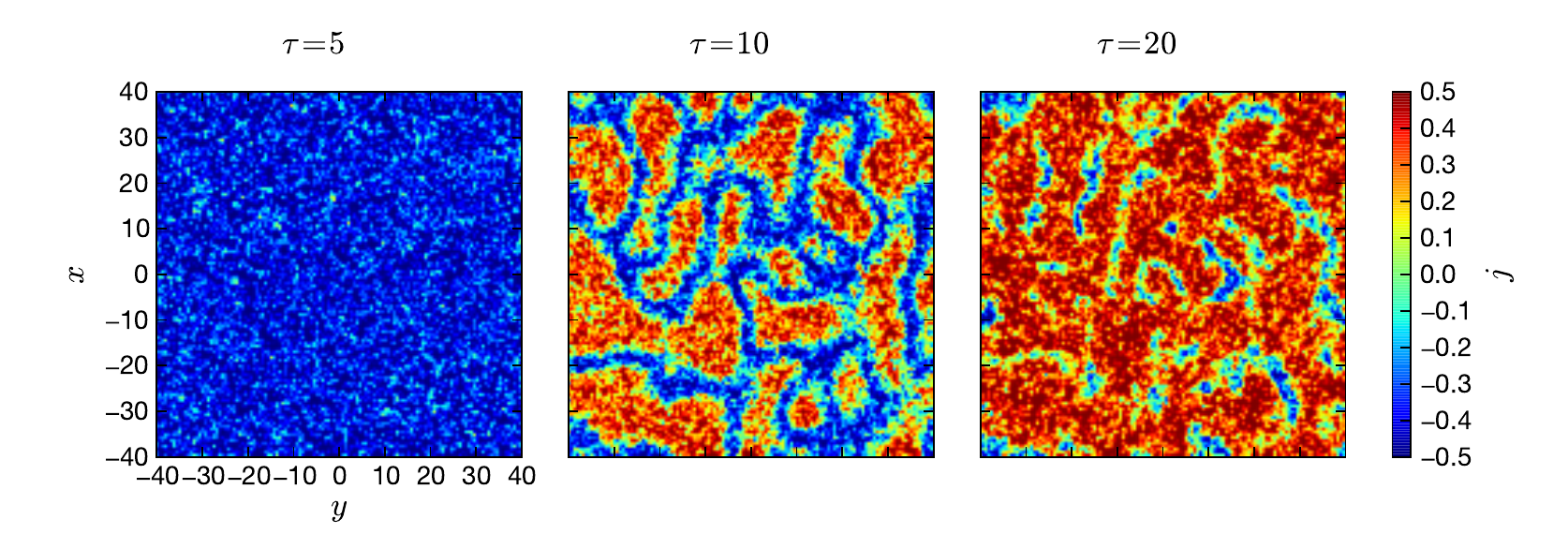}

\caption{\label{fig:Plot-of-3D singletrajectory}Plot of 3D spacial clustering
geometry of the local visibility $j$ for single trajectory, showing
contour plot vs space, sliced through a single z-coordinate. Coupling
$\tilde{\nu}=0.1$, condensate size $80\times80\times80$, $128\times128\times128$
spatial grid points, $\gamma=0.1$, 5000 time steps, opposite initial
phases. Plotted: a slice orthogonal to x-axis at $\tau=10$.}
\end{figure*}

Recent experiments have shown the possibility of exploring a condensate
in a ``can'' or cylindrical trap. This is a flat trap in three-dimensions,
with reflecting boundary conditions. While our simulations do not
have this type of boundary, it is feasible to carry out large three-dimensional
calculations relatively quickly using GPU hardware~\cite{Opanchuk2012}.
In Fig.~\ref{fig:Plot-of-3D singletrajectory}, we show a slice through
a condensate, in which one coordinate is held fixed, while the other
two are varied. This is a single ensemble member, and the other parameters
are similar to the two-dimensional case, Fig.~\ref{fig:Plot-of-2D singletrajectory}.

The resulting spatial geometry of clusters of hot and cold matter
is distinctly different to the two-dimensional case, with greater
connectedness, as larger coherent structures are formed over the same
time-scale. Here we have plotted the real spatial image of a two-dimensional
slice, to show domain-like clustering effects. This demonstrates that
very large-scale clustering effects occur even in the absence of explicit
gravitational interactions, for our model.

Rapid progress in imaging technology means that it is now possible
to image such 2D slices in the experiment~\cite{Bucker2009}, although
it is technologically more difficult than typical BEC experimental
techniques which produce a column density image of an expanded condensate.

\section{Summary}

The general behavior and unstable dynamics of an effective quantum
sine-Gordon field using a two-mode BEC with linear coupling has been
clearly demonstrated. Experimentally, for the scenario outlined here,
it is important to use an atomic species whose cross-coupling differs
from its self-coupling. This largely rules out the most commonly used
atomic species, $^{87}Rb$, which has very similar self and cross
couplings apart from a small region near a Feshbach resonance. A Feshbach-type
experiment is still possible, although losses are greater in this
regime. Apart from this case, there are many species which are both
able to be evaporatively cooled and have multiple hyperfine levels.
For example,$^{39}K$ can be used. It has a sizable difference of
$g_{1}g_{2}-g_{c}^{2}$~\cite{D'Errico2007,Chin2010}, and it is
therefore more suitable for realization of our proposal.

Our results will be extended to a more complete analysis of the numerical
simulations elsewhere, in which we will analyze in greater detail
the effects of different types of state preparation. However, it is
clear that in principle the use of ultra-cold quantum gases as quantum
simulators is not restricted to condensed matter and non-relativistic
analogies. Simulating interacting relativistic quantum fields in one,
two or three dimensions appears completely feasible, both experimentally
and computationally. An experimental implementation would throw much-needed
light on the question of how accurate our numerical approximations
are for these parameter values.

While we have focused on the practical issues of how to carry out
the experiment and corresponding numerical simulations, these issues
pale somewhat in comparison to another issue. We hope our proposal
will stimulate interest in this fundamental question, which is:
\begin{itemize}
\item \emph{how do we carry out measurements on the quantum universe, when
we are part of it?}
\end{itemize}

\section*{Acknowledgements}

The authors acknowledge helpful discussions with Andrei Sidorov, Tim
Langen and Uli Z\"ulicke. OF and JB were supported by the Marsden
Fund (contract No.~MAU0910). PDD was funded by the Australian Research
Council.

\bibliography{Qsim-universe}

\end{document}